Tuning macro-twinned domain sizes and the b-variants content of the adaptive 14-modulated martensite in epitaxial Ni-Mn-Ga films by co-sputtering

Jérémy Tillier[a, b, *], Daniel Bourgault[a], Philippe Odier[a], Luc Ortega[a], Sébastien Pairis[a], Olivier Fruchart[a], Nathalie Caillault[b], Laurent Carbone[b]

[a]Institut Néel/CRETA, CNRS et Université Joseph Fourier, BP166, F-38042 Grenoble Cedex 9, France

[b]Schneider Electric France, 38TEC/T1, F-38050 Grenoble Cedex 9, France

Abstract

In order to obtain modulated-martensite in our epitaxial Ni-Mn-Ga films, we have tuned the composition by using a co-sputtering process. Here we present how the composition affects the variant distribution of the 14-modulated martensite at room temperature. The nature of such modulated-martensites is still strongly debated for magnetic shape memory alloys. It has been very recently demonstrated that the modulated-martensites in Ni-Mn-Ga are adaptive phases. The results presented here corroborate this theory for the first time, for three different compositions. Moreover, we demonstrate with the help of the adaptive modulations theory that b-variants of the 14-modulated martensite form close to the free-surface of the film to release the stress induced by branching of macro-twinned domains during the martensitic transformation on a rigid substrate. At room temperature, the content of such b-variants is found to strongly decrease when the macro-twinned domain sizes increase.

1. Introduction

Shape Memory Alloys (SMA) exhibit a displacive phase transition from the high temperature austenite phase to a martensite phase with a lower crystallographic symmetry. The martensitic transformation (MT) requires the accommodation of the martensite on a habit plane. This lattice invariant interface fixes the geometrical relationship between both crystallographic structures, the lattice mismatch being relaxed by twinning in the martensite [1-4]. Such martensite structures are thus composed of twinned variants of different crystallographic orientations. The variants are separated by highly mobile twin boundaries, allowing rearrangement of the structure under rather low stress by preferential growing of the most favorable variant at the expense of the others. The maximal achievable strain of this super-plastic behavior depends on the tetragonality of the martensite unit cell and can reach up to 8% in Ni-Ti alloys [5].

Magnetic Shape Memory Alloys (MSMA) constitute a new class of SMA. In addition to the MT, these metallic alloys exhibit magnetic response. The magnetic properties of MSMA have been found to strongly depend on the alloy system. In Ni-Co-Mn-In alloys, the transformation from a cubic ferromagnetic austenite to a lower-symmetry non-magnetic or paramagnetic martensite gives rise to a large inverse magneto-caloric effect [6-7]. In Ni-Mn-Ga, the magneto-structural coupling between magnetic moments and martensite variants leads to a large panel of properties like Magnetic Induced Martensite (MIM) [8] or Magnetic Induced Rearrangement (MIR) of martensite variants [9]. This last effect has gained considerable attention as magnetic induced strains reaching 10% have been observed in Ni-Mn-Ga single crystals [10-16]. In fact, these strains are about two orders of magnitude larger than that commonly observed in magneto-strictive or piezo-electric materials [17].

Today large efforts are carried out to develop MIR-active Ni-Mn-Ga films because of their promising applications as new micro-actuators or micro-sensors for Micro-Electro-Mechanical Systems (MEMS) [3, 4, 9, 17 and 18]. Highest strains being only reported in bulk single crystals, epitaxial growth is considered to be the most hopeful process. Moreover, evidences of MIR in films have only been observed in epitaxially grown layers to date [4, 9, 17 and 18].

A key requirement to obtain MIR is that the magnetically induced stress exceeds the mechanical stress needed to rearrange the structure. Large magneto-crystalline anisotropy and low twinning-stress of the martensite phase are thus prerequisites. These conditions can only be fulfilled in modulated structures like the ten-modulated (10M) or the fourteen-modulated (14M) martensites. The twinning stress of the non-modulated (NM) martensite, which is thermodynamically the more stable phase, is too high to allow MIR [19].

At room temperature (RT), the structure of the martensite has been found to strongly depend on the alloy composition, the proportion of NM-martensite increasing with the average valence electron concentration per atom [19]. Tuning the composition of Ni-Mn-Ga films has been achieved by mean of various techniques like changing the target composition [3], the deposition temperature [4], the sputtering reactor pressure [20] or applying a negative bias voltage on the substrate [21].

In this article we use the simultaneous deposition of a ternary $Ni_{56}Mn_{22}Ga_{22}$ alloy and a pure manganese target in order to tune the film composition. All parameters have been fixed to facilitate epitaxial growth on (001) MgO substrates and the composition has been varied by increasing the applied power on the manganese target. In all of our films, austenite, 14M-martensite and NM-martensite coexist at room temperature, due to too high (e/a) factors. Here we demonstrate that the variant distribution of the 14M-martensite depends on the composition. Orientations of each phase have been determined by pole figures measurements. The results corroborate for the first time the theory of adaptive modulations of Ni-Mn-Ga martensite, which was very recently developed by Kaufmann et al. [2] following the Khachaturyan's concept [1]. Finally, the study focuses on the influence of composition on the surface morphology of twins. The theory of adaptive martensite and normalized integrated intensities of pole figures are also used to study the content evolution of 14M b-variants.

2. Experimental

Epitaxial Ni-Mn-Ga films have been deposited by magnetron sputtering. A low residual pressure in the range of $10^{-6}$ Pa was used to avoid any oxidation of the films. Epitaxial films have been grown in a confocal sputtering reactor equipped with six cathodes: three operated in Direct Current (DC), three operated at Radio Frequency (RF). In the following, we use only two of these. The ternary $Ni_{56}Mn_{22}Ga_{22}$ alloy was inserted in a DC-cathode whereas the pure Mn target was inserted in a RF-cathode to enable tuning of the composition. In order to ensure chemical homogeneity and constant film thicknesses, the depositions were made with the substrate rotating at a speed of 5 rpm. The (001) MgO monocrystalline substrate temperature has been fixed at 773 K and the applied power (86 W) on the ternary target was optimized to obtain a deposition rate of 1 $\mu m.h^{-1}$. The sample batch has been deposited by increasing the power applied on the Mn target from 0 to 30W for a fixed deposition time of a half hour. The deposition rate of the Mn target was $(2/3).10^{-2}$ µm.h$^{-1}$W$^{-1}$, leading to film thicknesses ranging from 500 nm (no power on the Mn target) to 600 nm (30W on the manganese target). More details on the deposition process can be found in [22].

The film compositions were determined by Energy Dispersive X-ray (EDX) spectroscopy using a JEOL 840A Scanning Electron Microscope (SEM). The structural and textural characterizations were realized using X-Ray Diffraction (XRD). Measurements have been carried in a four-circle instrument (Seifert MZ IV) with the copper K$_\alpha$ radiation. The diffractometer was equipped with an optic (Xenocs) enabling a low divergence (0.06°) of the X-ray beam and a rear mono-chromator in order to enhance the signal to noise ratio. Alignment of the samples has been realized using the (002) reflection of the MgO substrate. It allows probing crystallographic orientations of all Ni-Mn-Ga phases in absolute coordinates, the MgO substrate being used as a reference system [2]. θ-2θ scans have been measured for tilt angles ψ ranging from zero to ten degrees, at two rotation angle Φ selected with respect to the epitaxial relationship between substrate and austenite, in order to determine optimized 2θ positions of the (400) (040) (004) martensites tilted lattice planes. After that, pole figures of (400) martensite reflections have been acquired in the range Φ 0 to 360° and ψ 0 to 10° for both the NM-martensite and the 14M-martensite. The microstructures were analyzed using various microscopy techniques. Optical microscopy with polarized light (Zeiss microscope equipped with a Hamamatsu ORCA-ER digital camera) and Back-Scattered Electron (BSE) detectors of a Field-Emission Scanning Electron Microscope (FESEM Zeiss Ultra +: 8kV, 15mm, 100X) have been used to investigate sizes and morphologies of the macro-twinned domains at the surface of the films.

3. Results and discussion

3.1. Tuning the composition

As demonstrated in reference [22], the multi-target sputtering reactor allows an independent control of the applied power on each target. The flux of sputtered manganese atoms increases with the power applied on the manganese target. It leads to rising up the manganese atomic content whereas proportions of nickel and gallium decline. The film compositions have been varied from $Ni_{60}Mn_{20}Ga_{20}$ (no power on the manganese target) to $Ni_{48}Mn_{36}Ga_{16}$ (30W on the manganese target). In this study we have selected three samples: NMG1, NMG2 and NMG3. These epitaxial films have been deposited for powers applied on the manganese target of 15, 20 and 25W, respectively. Table 1 indicates the elemental compositions and the average valence electron concentration per atom (e/a) of the selected samples, where the (e/a) factors have been calculated as follow:

$$(e/a) = \frac{10.Ni_{at.\%} + 7.Mn_{at.\%} + 3.Ga_{at.\%}}{100} \quad (1)$$

3.2 Textural investigations and adaptive modulations of the 14M-martensite

The martensitic transformation from the parent austenite phase epitaxially grown during deposition at high temperature to the martensite phase must accommodate the strain induced by the crystal lattice mismatch between both phases. The transformation path requires an invariant habit plane along which each cell rotates to restore crystal-lattice continuity across the boundary plane [1]. In order to identify the martensite phases and determine their orientations relative to the substrate, θ-2θ XRD-scans were measured in the 2θ-range of (400) reflections of Ni-Mn-Ga martensites for tilt angles $\psi$ ranging from 0° to 10° in a preliminary work [22]. The rotation angles $\Phi$ were selected with respect to the A/MgO epitaxial relationship and twinning planes of Ni-Mn-Ga martensites, which are of (101) type for the bct cell [22]. All XRD-scans series have revealed the coexistence of three phases, i.e. austenite, 14M-martensite and NM-martensite that coexist at room temperature in our films, even for the $Ni_{60}Mn_{20}Ga_{20}$ composition, which lead to an e/a factor of 8.0 [22].

Thomas et al. [4] demonstrates that in the case of Ni-Mn-Ga films epitaxially grown on MgO substrates, a thin interfacial austenitic layer persists on the film/substrate interface, even at temperatures below the martensitic finish temperature of the film volume. The presence of this austenite layer is due to substrate-induced constrains which impede the martensitic transformation at the substrate interface [4]. The presence of 14M-martensite even for compositions that lead normally to NM bulk alloys is of particular interest. Due to the crystallographic requirements, the NM cells cannot accommodate on the remaining austenite. As discussed in reference [4], accommodation of the martensite on the austenite thin layer occurs by twinning of the 14M-martensite, which possesses a crystallographic axes b equal to the cubic cell parameter of austenite. Only four of the six possible variants can exist at the austenite/martensite interface [4]. The two variants with b axis of the 14M-martensite pointing out-of-plane, which are denominated primary b-variants in the following, are not allowed by crystallography at the austenite/martensite interface [4]. Kaufmann et al. have recently described the concept of the adaptive nature of Ni-Mn-Ga modulated martensites [2]. This concept was first discussed by Khachaturyan et al. [1], which demonstrated that modulated martensites of Ni-Al and Fe-Pd alloys were adaptive phases, composed of nanoscopic variants of tetragonal building-blocks. Kaufmann et al. have shown that the 14M-martensite of the Ni-Mn-Ga system is an adaptive phase constructed with tetragonal building-blocks of the NM-martensite.

The cell parameters of each phase, determined from $\psi$-dependent $\theta$-$2\theta$ XRD-scans are presented in table 2. Table 2 also shows the 14M cell parameters predicted by the adaptive modulations theory: $a_{14M}=c_{NM}+a_{NM}-a_A$, $b_{14M}=a_A$ and $c_{14M}=a_{NM}$ [1 and 2]. Despite of the fact that the measured tetragonalities of 14M are lower than that predicted from adaptive theory, the key precondition for a coherent austenite/14M interface, i.e. $b_{14M}=a_A$ is fulfilled for all the films, as reported in the study of Kaufmann et al. [2]. The adaptive theory also allows to prognosticate the stacking sequence of the adaptive 14M, as $d_1/d_2 = (a_A-a_{NM})/(c_{NM}-a_A)$ [2]. Table 2 shows that the periodicities expected from the adaptive theory are close to the ideal value of 2/5=0.4, describing a perfect $(5\underline{2})_2$ stacking order [2]. The changes in $d_1/d_2$ suggest occurrence of stacking faults, which can explained the observed difference between predicted and measured cell parameters [2].

The (400) pole figures of NMG1 and NMG3 are presented in figure 1, both for the 14M and the NM-martensite. Each pole figure exhibits four-fold symmetry, indicating that epitaxial growth occurs for all the samples [2-4, 9]. No other reflections than those shown on this figure were observed. The discussion will now focus on the pole positions, according to WLR and adaptive modulations theories. Twinning of the 14M-martensite on the austenite layer is described by the WLR theory, neglecting the monoclinic distortion of the pseudo-orthorhombic cell, as $\psi_{14M}=\pi/4-\text{atan}(c_{14M}/a_{14M})$ [4]. Therefore (400) and (004) reflections of the 14M-martensite contribute at a same $\psi$ angle. The difference between the simplified model and observed peak positions is due to the monoclinicity of 14M, which is neglected in this simplified model [4]. The peak at the center of 14M(040) poles figures originates from the residual austenite layer and, possibly, some primary b-variants of the 14M which cannot exist close to the substrate [4]. Nevertheless, as it will be demonstrated in the following, primary 14M b-variants may be found near the free-surface of the film, in order to release the transformation stress of the substrate-constrained film. Additional peaks on (040) and (004) pole figures of the 14M martensite arise from secondary generation of twin along the film thickness [4]. Note that all pole figure exhibits low intensities and high background. In fact, only primary and secondary variants contribute to the observed poles [4]. Further generations of twins lead to increase intensity of the background. So that, the entire peak positions of 14M pole figures are explained. Understanding NM pole positions requires the use of the adaptive theory [2]. According to reference [2], macroscopic NM variants adapt on the NM nano-twinned variants constituting the 14M super-cell by branching. Therefore, no changes of NM variants orientations between nanoscopic and macroscopic twins are expected [2]. From the unit cell parameters of NM building-blocks and assuming a $(5\underline{2})_2$ stacking sequence for the 14M-martensite, the angles $\beta_1$ and $\beta_2$ between b-axis of the 14M super-cell and, respectively, c and a-variants of NM building-blocks can be determine from basic geometry. A 3D model of the 14M super-cell constructed from NM building-blocks is available in [2]. The $\beta_1$ and $\beta_2$ angles can be calculated as follows, the angle $\alpha$ between nano-twinned NM blocks being given by WLR theory:

$$\beta_1 = \alpha - \arctan\left[\frac{3}{7}\tan(\alpha)\right] \text{ and } \beta_2 = \alpha + \arctan\left[\frac{3}{7}\tan(\alpha)\right] \text{ with } \alpha = \frac{\pi}{4} - \arctan\left[\frac{a_{NM}}{c_{NM}}\right] \quad (2)$$

To determine the NM pole positions, tilts of NM variants have to be applied for each 14M variants. It can be realized by using unity vectors and rotation matrices and then transforming to spherical coordinates to obtain $\psi$ and $\Phi$ angle of the pole figures (see [2] for more details).

Figure 2 presents both the measured peak positions and that calculated from adaptive theory for the three selected samples. Figure 2 reveals the good agreement between experiment and model. The experimental results presented here thus corroborate the adaptive theory of modulated martensites recently presented in reference [2], for three different compositions. The difference between experiment and model can be explained because of the slight monoclinic distortion of the 14M-martensite which is neglected to calculate the tilt angle between residual austenite or substrate and 14M-martensite [2, 4].

The adaptive theory also predicts the presence or absence of the different 14M variants. The NM (004) pole figure of NMG1 shown on figure 1 demonstrates reflections which can only originate from primary b-variant of the 14M. These peaks at tilt angle around 8.5° are not observed for NMG3 (see the areas surrounded in red on figure 1). Thus, pole figure measurements of the NM-martensite combined to the use of the adaptive modulations theory suggest that primary 14M b-variants do not exist for this last sample.

3.3   Surface macro-twinned domain sizes and proportion evolution of  primary 14M b-variants

Figure 3 highlights morphologies of the surface macro-twinned domains. The macro-twinned domains are revealed by optical microscopy with polarized light, figure 3.a., b., and c. respectively corresponding to NMG1, NMG2 and NMG3. The strongly contrasted areas, which are clearly seen in the figures, belong to macro-twinned domains with two different crystallographic orientations. It should be noted that branching between two macro-twinned domains occurs in MgO [110] type directions.  Figure 3.a, 3.b and 3.c clearly bring to light that for decreasing (e/a) factors, the macro-twinned domain sizes increase. FESEM observations realized in BSE mode (Figure 3.d., e. and f. for respectively NMG1, NMG2 and NMG3) reveal parallel lines in two perpendicular directions. The directions of observed lines are along MgO [100] and MgO [010]. These lines, with a clearer BSE contrast, correspond to non-twinned areas at the film surface level. It is demonstrated on figure 4, which shows a SEM image of NMG1, realized by acquiring a mixed signal of both SE and BSE detectors of the FESEM. The figure reveals that along the [100] or [010] MgO directions, the macro-twinned domains are sometimes separated by non-twinned areas at the surface level. In order to acquire a mixed signal of both SE and BSE detectors, the high voltage and the working distance are optimized for the BSE detector, explaining why the resolution of the SE information is so low.

We will now discuss the origin of the non-twinned areas at the surface level. The BSE contrast images (Figure 3.d. and f.) bring to light that a lot of non-twinned areas are observed on the surface of NMG1 whereas NMG3 exhibit a very low content of non-twinned areas. The same evolution is observed on the NM (004) pole figures. In fact, the NM (004) pole figure of NMG1 exhibits splitted peaks at tilt angles around 8.5°, whereas these peaks are not detected for NMG3 (see the areas surrounded in red on figure 1). From the adaptive modulations theory, these peaks can only originate from primary 14M b-variant, with their b-axis pointing out-of-plane. As discussed above, these variants cannot exist at the austenite/martensite interface but are expected to form close to the free-surface to release the transformation stress [4]. It is thus reasonable to believe that the non-twinned areas at the film surfaces correspond to the primary14M b-variants. These variants are twinned in the direction of the film thickness, explaining why these areas appeared non-twinned when investigating the film surface.

The qualitative evolution of the relative content of each 14M-martensite variants can be estimated by using integrated intensities of the pole figures. For the relative content $F_R$ of both primary 14M b-variants and austenite, it can be obtained by calculating the normalized integrated intensity at the low-angle range of the (040)14M pole figure as follow:

$$F_R \text{ (primary 14M b-variants + austenite)} = \frac{\int_{\Phi=0°}^{\Phi=360°}\int_{\Psi=0°}^{\Psi=2°} I_{(040)}^{14M} \partial\Phi\, \partial\Psi}{\sum_{(400),(040),(004)}^{14M} \int_{\Phi=0°}^{\Phi=360°}\int_{\Psi=0°}^{\Psi=10°} I \partial\Phi\, \partial\Psi} \quad (3)$$

The results, which are depicted on figure 5, only give a qualitative idea on the content evolution of primary 14M b-variants. At first glance, it is not possible to distinguish the contribution of the remaining austenite and that of the primary 14M b-variants, because this two reflections overlap at the center of the (040) 14M pole figure. Nevertheless, the observation of BSE images of figure 3, correlated with the disappearance of peaks on the (004) NM pole figure, has demonstrate that sample NMG3 exhibit a very low content of primary14M b-variants. The content of around 20% found for this sample, which possesses the lowest (e/a) factor, thus arise from the remaining austenite. For sample NMG3, the austenite contributes for around 20% of the intensity diffracted in the 14M pole figures. Assuming a constant contribution of this thin interfacial austenite, figure 5 shows that the relative content of primary 14M b-variants increases when the (e/a) factor increases, while the content of the other variants decreases. For sample NMG1, which show an important proportion of lines with clearer contrast on the BSE image, the content increases to around 55%, indicating that the film show a high content of primary 14M b-variants. This result confirm the presence of intense peaks at tilt angle around 8.5° for the (004) NM pole figure of NMG1 as well as the large amount of clearer lines on the BSE image of figure 3.d, these observations being indications of 14M variants with their b-axis pointing out-of-plane, at the free-surface of the film.

4. Conclusions

A batch of three epitaxial Ni-Mn-Ga films of different compositions has been studied. All of our films demonstrate the coexistence of interfacial austenite, 14M-martensite and NM-martensite. The orientations of each phase have been determined by pole figure measurements. The experiments agree with the pole positions calculated from WLR and adaptive modulations theories. These results thus corroborate the very recent theory of adaptive modulations of Ni-Mn-Ga martensites. The surface morphologies of twinned areas are also strongly affected by the composition, the size of macro-twinned domains increasing when the average valence electron concentration is decreased. Moreover, non-twinned areas have been observed at the free-surface of the films. The use of the adaptive modulations theory as well as normalized integrated intensities of pole figures has confirmed that these non-twinned areas belong to 14M variants with their b-axis pointing out-of-plane. The proportion of such 14M b-variants strongly decreases when the size of macro-twinned domains increases. It suggests that these variants contribute to the relaxation of the stress induced by branching of macro-twinned domains during the martensitic transformation on a rigid substrate.

Figure captions

Figure 1:

NMG1 and NMG3 pole figures of (400), (040) and (004) reflections for both the 14M and the NM martensites. A schematic of the representation of pole figures with the ranges of the tilt angle $\psi$ and the rotation angle $\Phi$ is given on the figure for more clarity.

Figure 2:

Comparison of NM pole positions determined from experiment (open symbols) and from WLR and adaptive modulations theories (closed symbols) for NMG1, NMG2 and NMG3. The pole positions of the NM (400) and the NM (004) reflections are represented with circles and squares, respectively. Only one quadrant of the pole figure is shown for each sample.

Figure 3:

Surface-morphologies of macro-twinned domains revealed both by polarized light (a., b. and c.) and by FESEM in BSE mode (d., e. and f.).

Figure 4:

Surface-morphology of NMG1, characterized by acquiring a mix signal of both the SE and BSE detectors of the FESEM. The image demonstrates that the clearer lines of BSE images belong to non-twinned areas at the surface level.

Figure 5:

Qualitative evolution of the relative content of the primary 14M b-variants and austenite compared to that of all other variants. The BSE images of each film, which are the same as in figure 3, are depicted on the figure for a better understanding.

Table captions

Table 1:

Compositions and average valence electron concentrations of the selected epitaxial Ni-Mn-Ga films.

Table 2:

Experimental cell parameters of the 14M-martensite, the NM-martensite and the austenite for the three selected epitaxial films as well as the 14M-martensite lattice parameters predicted by the adaptive modulations theory from cell parameters of austenite and NM-martensite determined from Bragg reflections.

Table 1

| Sample | RF-power on Mn [W] | Ni [at.%] | Mn [at.%] | Ga [at.%] | e/a [e-/at.] |
|---|---|---|---|---|---|
| NMG1 | 15 | 54.6 | 27.4 | 18.0 | 7.92 |
| NMG2 | 20 | 52.8 | 30.1 | 17.1 | 7.90 |
| NMG3 | 25 | 49.9 | 33.8 | 16.3 | 7.85 |

Table 2

| | Experimental | | | | | | Model | | | |
|---|---|---|---|---|---|---|---|---|---|---|
| | 14M | | | NM | | A | 14M | | | |
| | a [Å] | b [Å] | c [Å] | a [Å] | c [Å] | a [Å] | a [Å] | b [Å] | c [Å] | d1/d2 |
| NMG1 | 6.31 | 5.81 | 5.63 | 5.39 | 6.71 | 5.81 | 6.29 | 5.81 | 5.39 | 0.47 |
| NMG2 | 6.32 | 5.81 | 5.65 | 5.40 | 6.73 | 5.81 | 6.32 | 5.81 | 5.40 | 0.45 |
| NMG3 | 6.31 | 5.83 | 5.61 | 5.43 | 6.78 | 5.83 | 6.38 | 5.83 | 5.43 | 0.42 |

Figure 1

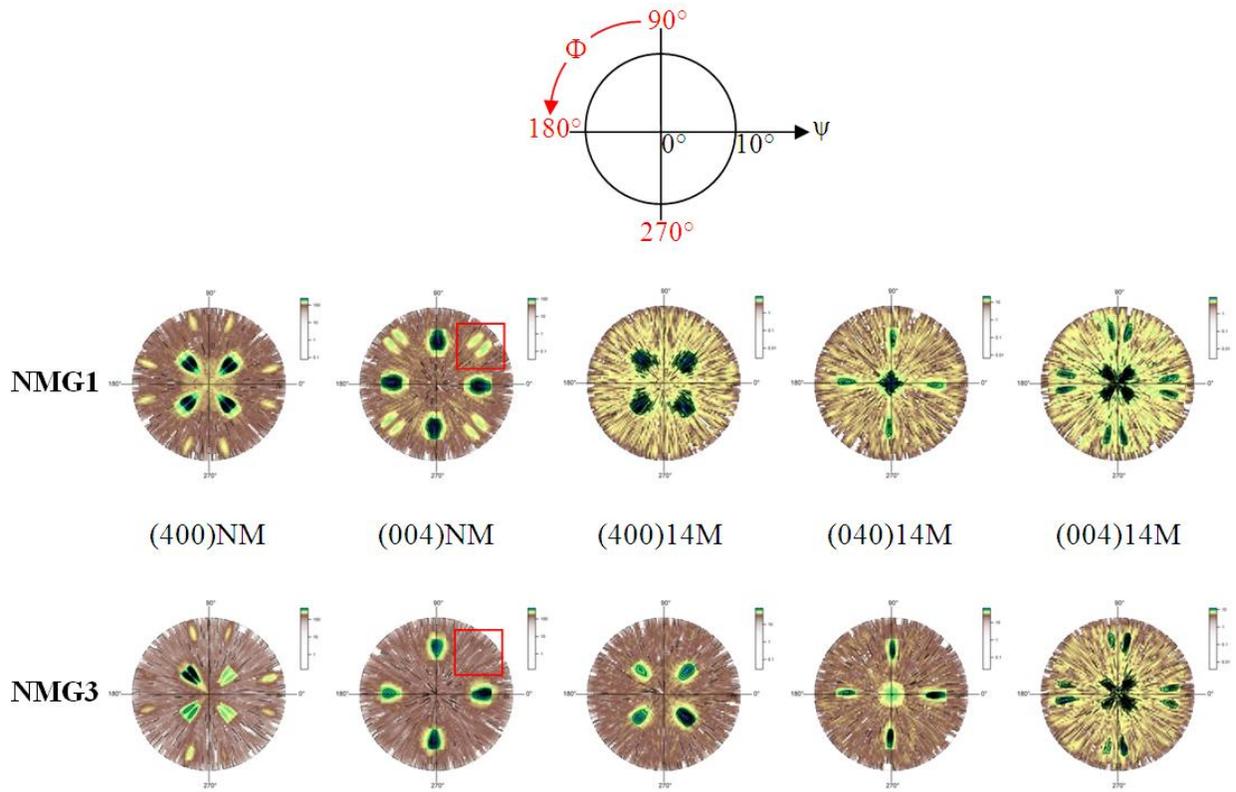

Figure 2

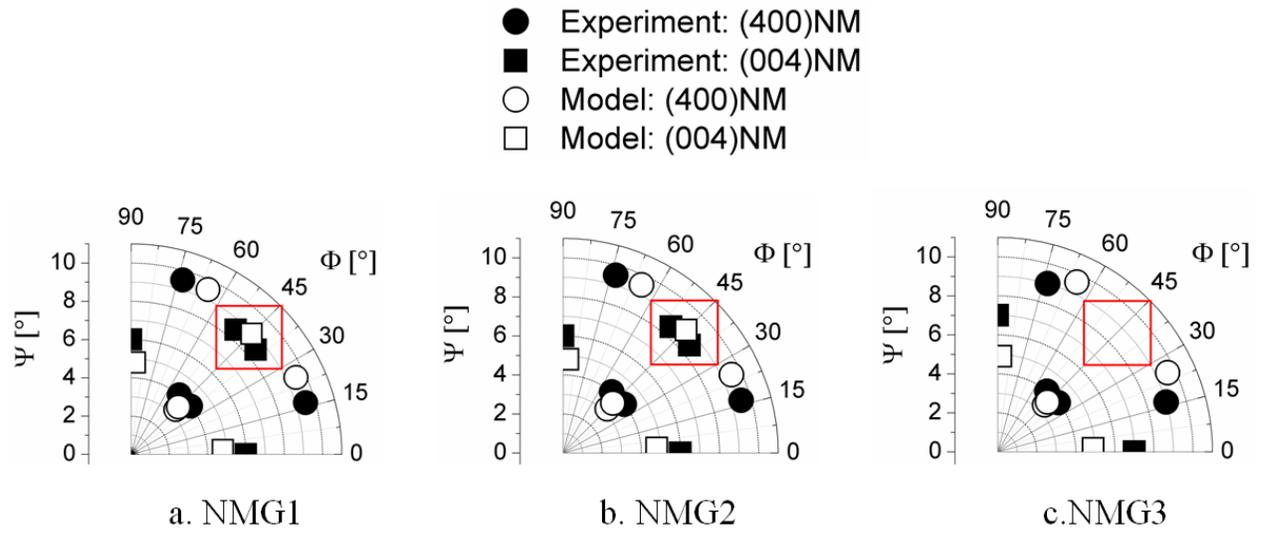

a. NMG1  b. NMG2  c. NMG3

Figure 3

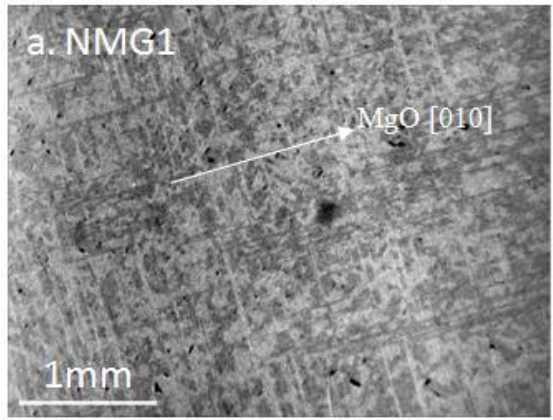
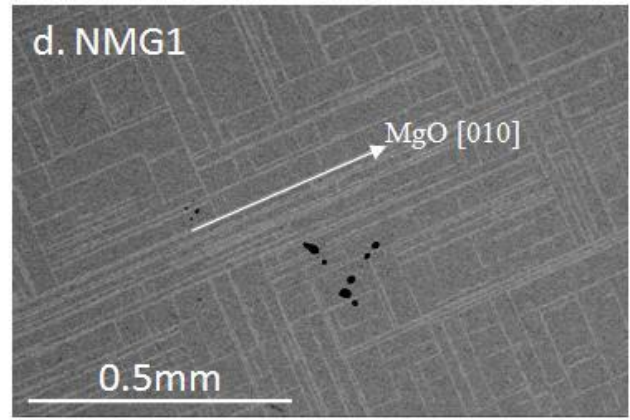
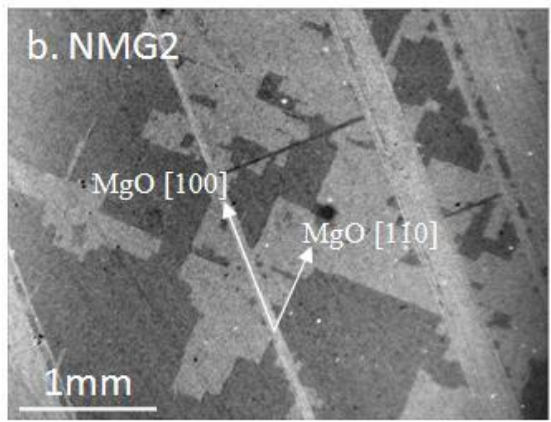
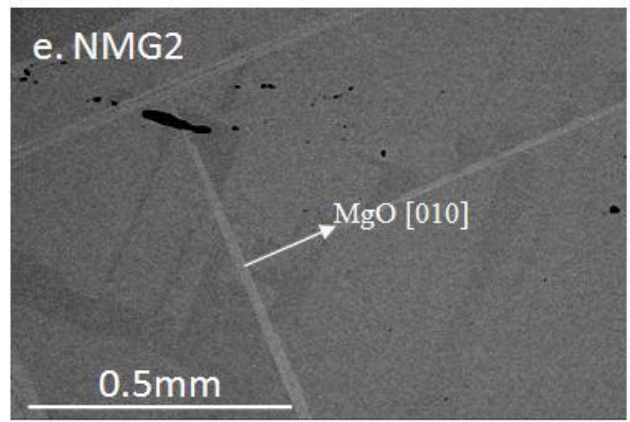
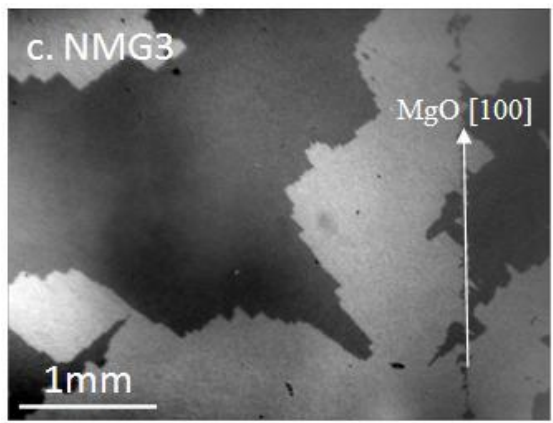
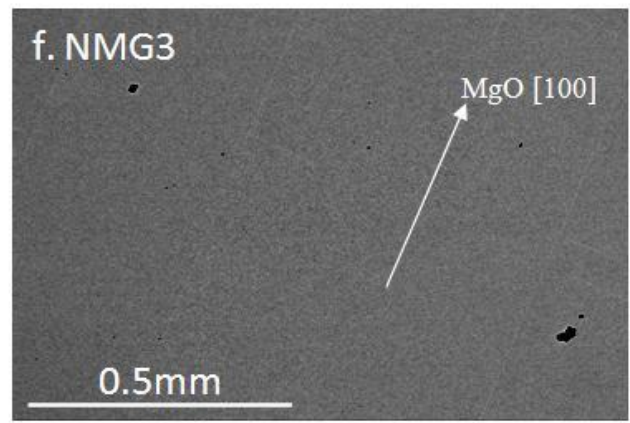

Figure 4

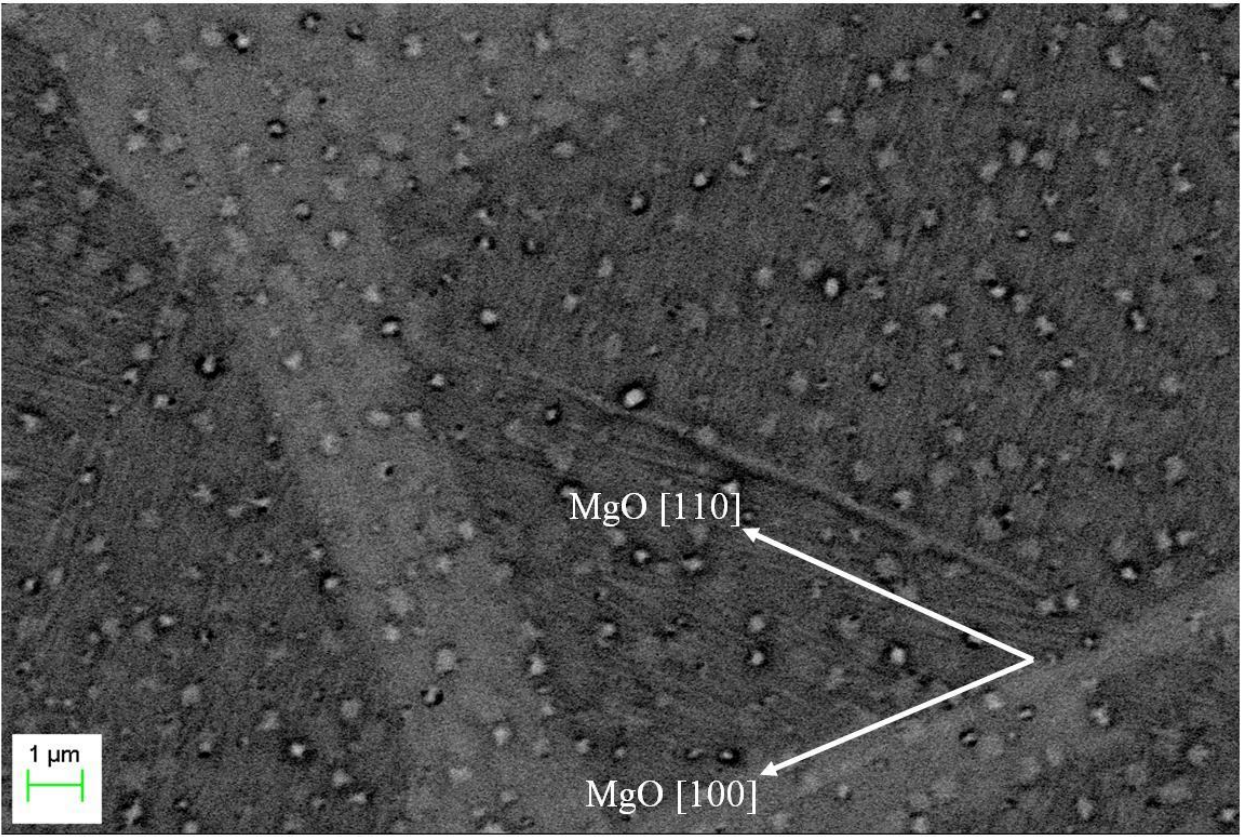

Figure 5

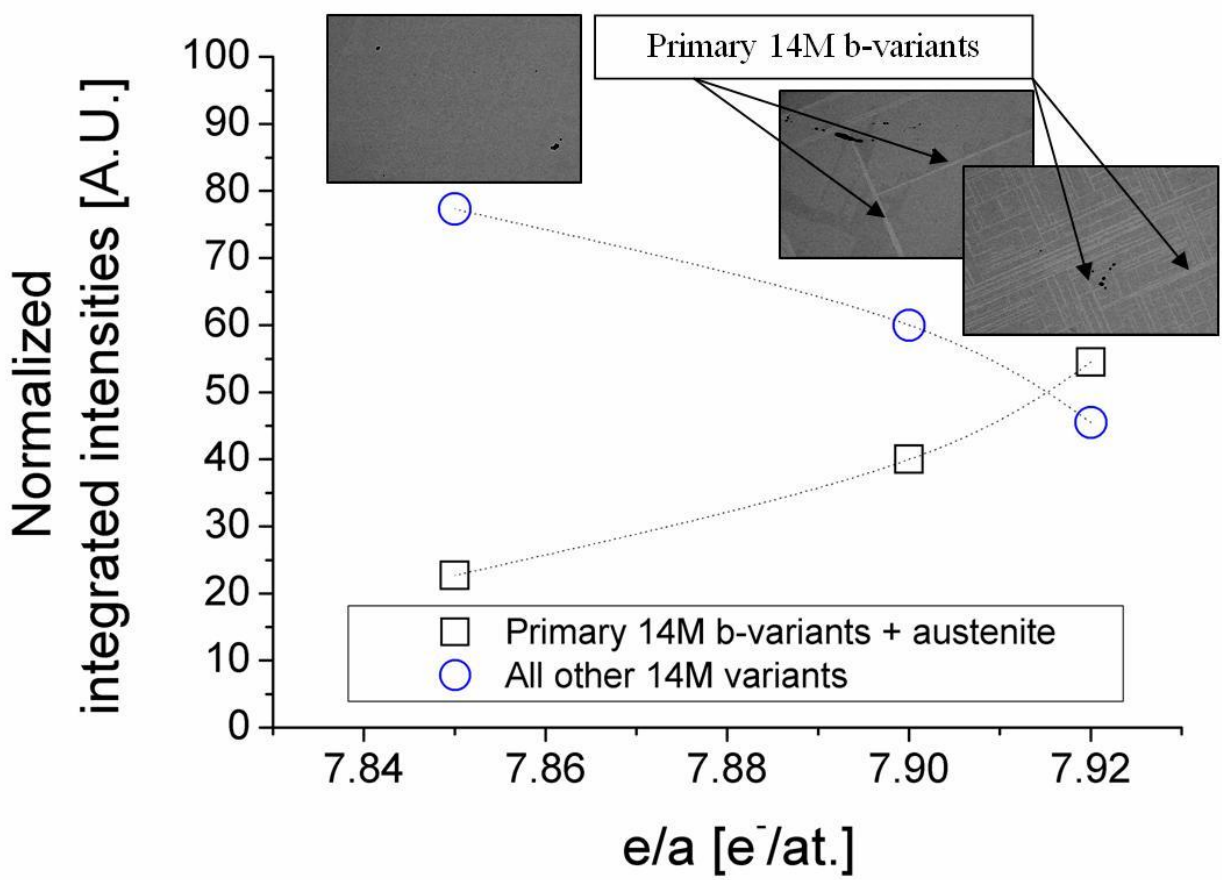